\documentclass[conference]{IEEEtran}
\IEEEoverridecommandlockouts
\usepackage{cite}

\ifCLASSINFOpdf
\else
\fi
\usepackage{algorithmic}
\usepackage{algorithm}
\usepackage{graphicx}
\usepackage{caption}
\usepackage[utf8]{inputenc}
\usepackage[english]{babel}
\usepackage{epsfig}
\usepackage{amsfonts}
\usepackage{ifpdf}
\usepackage{multirow}
\usepackage{amssymb}
\usepackage{amsthm}
\usepackage{mathtools}
\usepackage{mathrsfs}
\usepackage{textgreek}
\usepackage{grffile}

\newcommand{\norm}[1]{\left\lVert#1\right\rVert}
\DeclareMathOperator{\sign}{sgn}

\DeclareMathOperator*{\argmax}{arg\,max}

\usepackage[usestackEOL]{stackengine}
\def\equA{\mathrel{\stackon[1.8pt]{=}%
  {\scriptstyle (a)}}}

\stackMath

\ifCLASSOPTIONcompsoc
 \usepackage[caption=false,font=normalsize,labelfont=sf,textfont=sf]{subfig}
\else
  \usepackage[caption=false,font=footnotesize]{subfig}
\fi
\begin{document}
%
\title{Differential Deep Detection in Massive MIMO With One-Bit ADC}



%
\author{\IEEEauthorblockN{Don-Roberts Emenonye\IEEEauthorrefmark{2},
Carl Dietrich\IEEEauthorrefmark{2}, and
R. Michael Buehrer\IEEEauthorrefmark{2}
}
\thanks{\IEEEauthorrefmark{2} Don-Roberts Emenonye, Carl Dietrich, and R. Michael Buehrer are with  Wireless@VT, Bradley Department of ECE,
Virginia Tech, Blacksburg, VA, 24061. (Emails: \{donroberts, cdietric, rbuehrer\}@vt.edu)}

}


\maketitle

\begin{abstract}
This article presents a differential detection scheme for the uplink of a  massive MIMO system that employs one-bit quantizers on each receive antenna. We focus on the detection of differential amplitude and phase shift keying symbols and we use the Bussgang theorem to express the quantized received signal in terms of quantized signals received during previous channel uses. Subsequently, we derive the maximum likelihood detector for the differentially encoded amplitude and phase information symbols. We note that while the one-bit detector can decode the differentially encoded phase information symbols, it fails to decode the differentially encoded amplitude information. To decode the amplitude information, we present a one-bit variable quantization level (VQL) system and train a deep neural network to perform two-symbol differential amplitude detection. Through Monte-Carlo simulations, we empirically validate the performance of the proposed amplitude and phase detectors. The presented numerical results show that the spectral efficiency attained in one-bit differential systems is better than the spectral efficiency attained in one-bit coherent systems.

\end{abstract}



%
\IEEEpeerreviewmaketitle

\section{Introduction}
The advantage of massive multi-input-multi-output (MIMO) lies in its ability to provide large multiplexing and diversity gains \cite{6736761}. However, to achieve these gains, some challenges need to be resolved. One of these challenges lies in the prohibitively high circuit power consumed by a large number of radio-frequency (RF) chains at the base station (BS). A point of power inefficiency in the RF chains occurs at the analog-to-digital converters (ADCs). In particular, it has been shown that an increase in the resolution of these converters causes an exponential increase in their power consumption \cite{761034}. To tackle this problem, several authors have investigated the operation of massive MIMO systems in the low-resolution regime.

In \cite{7088639}, a quantized distributed reception scheme is investigated for use in the downlink of a multi-antenna system that is serving a large number of internet-of-things (IoT) devices. In \cite{7439790}, the problem is reformulated for the uplink and a near maximum likelihood detector is presented. In \cite{7458830}, the problem is extended to MIMO systems with orthogonal frequency-division multiplexing (MIMO-OFDM) and a maximum a-posterior (MAP) algorithm for symbol detection is investigated. In \cite{8322194}, a downlink massive MIMO system with a relay is investigated. The relay system comprises one-bit analog-to-digital and one-bit digital-to-analog converters. In that work, the Bussgang decomposition \cite{9307295} is employed to develop a channel estimation technique. Although the power consumption is reduced by the application of one-bit ADCs, this perceived gain comes at the cost of a more expensive channel estimation phase.  In particular, channel estimation in the low-resolution regime employ longer pilot sequences \cite{7088639,7439790,7094619}. Hence, the channel is required to stay constant for a longer period of time. Such long pilot sequences negatively impact the spectral efficiency of coherent one-bit system.

\emph{To solve this challenge, we present an analysis of the uplink of a system with differential modulation employed at a single user transmitter and one-bit ADCs at the base station}. In general, differential/non-coherent communication is a well developed area \cite{275303,681323,340470,4024308} that has received considerable interest in recent times. In  \cite{5628284,4024308}, authors develop a look-up table based differential modulation technique. The table is constructed by minimizing the non-coherent distance between distinguishable codewords. In \cite{6780655}, an autocorrelation-based decision-feedback scheme is adopted for differential detection. In \cite{6814078,7037381}, the minimum non-coherent distance is analyzed and used to design constellations for non-coherent modulation.   \emph{In all these prior works, the impact of low-resolution ADCs on differential modulation is never considered, hence, in this article, we aim to develop detectors for differential detection in the low-resolution regime. While we employ an approximate maximum likelihood detector for differential phase detection, we employ a neural network (NN) for amplitude detection}.

In the existing literature, the plethora of data available either through simulation or real-world measurements has enabled researchers to explore the possibility of a completely data-driven communication model. This method involves approximating the submodular blocks at the transmitter and at the receiver with two disjoint functions. Neural networks, as general and well-known function approximators, make this a viable research option. Leveraging this, a novel end-to-end deep learning-based communication system was developed with both the transmitters and the receivers replaced by the encoder and decoder of an autoencoder \cite{8054694}. In that work, the parameters of the transmitter and the receiver are jointly updated by optimizing a loss function during training. In \cite{8262721}, the deep learning-based communication system was extended for multi-antenna transmission. More recently, deep learning has been used to enable energy-based detection in non-coherent systems. In \cite{9036067}, an end-to-end non-coherent learning-based system is developed. 

\section{SYSTEM MODEL}
In this article, we consider a single user employing a differential amplitude phase shift keying (DAPSK) system. The encoding scheme comprises of two concentric circles with different radius (amplitude). We define the amplitude of the inner circle as $\psi_0$, while $\psi_1$ represents the amplitude of the outer circle. The ring ratio is defined as $a = \frac{\psi_1}{\psi_0}$. A unit power constraint is enforced by $\psi_0^2 + \psi_1^2 = 2$. The collection of all points on the inner and outer circles form two phase shift constellations defined as  $\mathcal{S}_1 = \{x^{(1)}, x^{(2)}, \cdots x^{(M)} \}$, and $\mathcal{S}_2 = \{x^{(M+1)}, x^{(M+2)}, \cdots x^{(2M)} \}$ respectively. Both circles define the following constellation set, $\mathcal{S} = \{ \mathcal{S}_1, \mathcal{S}_2 \}$. $M$ represents the number of points on a particular circle. At the $v$th channel use, the transmitted symbol $x[v]$ depends on a block of $N_b$ bits defined as $\boldsymbol{b}[v] = [b_{1}[v], b_{2}[v], b_{3}[v],\cdots, b_{N_b}[v]]$. The last $N_b - 1$ bits in the block specify the phase of the transmitted symbol, while the first bit $b_{1}[v]$ specify its amplitude. If we define $\Upsilon$ as a function that generates phase shift keying symbols from a block of bits, then the phase information symbol can be generated as $s[v] = \Upsilon([ b_{2}[v], b_{3}[v],\cdots, b_{N_b}[v]] )$.
\begin{equation}
\label{equ:se_II_16APSK_encoding}
\begin{aligned}
a[0] =& \psi_{0}, c[0] = a[0],\\
c[v] =& c[v - 1] s[v], \\
x[v] =& a[v] c[v], \\
a[v] =&   \begin{cases}
      1 , & \text{if $b_{1}[v] = 0$},\\
     \frac{\psi_1}{\psi_0} ,  & \text{if $b_{1}[v] = 1$ and $\Tilde{x}[v - 1] = \psi_0$ },\\
     \frac{\psi_0}{\psi_1}  ,  & \text{if $b_{1}[v] = 1$ and $\Tilde{x}[v - 1] = \psi_1$ },\\
    \end{cases} \\
\end{aligned}
\end{equation}
where $a[v] \in \mathcal{A}  \{1, \frac{\psi_0}{\psi_1}, \frac{\psi_1}{\psi_0}\}$, $\Tilde{x}[v] = |x[v]|$, and $\Tilde{x}[v] \in \mathcal{X}\{\psi_0, \psi_1\}$. Clearly, the transmitted symbol  switches between $\mathcal{S}_1$ and $\mathcal{S}_2$, if $b_{1}[v] = 1$.
During the $v-$th channel use, the signal received at the base station can be represented as
\begin{equation}
\label{equ:section_II_received_signal}
\boldsymbol{y}_{u}[v] = \boldsymbol{h}[v]x_[v] + \boldsymbol{z}[v],
\end{equation}
where $\boldsymbol{y}[v] =  [{y}_{1}[v], {y}_{2}[v], \cdots, {y}_{U}[v] ]$, $\boldsymbol{h}[v] =  [{h}_{1}[v], {h}_{2}[v], \cdots, {h}_{U}[v] ]$ , and $\boldsymbol{n}[v]$ is the additive noise vector with $\boldsymbol{z}[v] \sim \mathcal{C}\mathcal{N}(0,\sigma^2\boldsymbol{I})$.  The frequency-selective channel at the $v-$th channel use can be modeled as a combination of $l$ parallel frequency-flat subchannels expressed as
\begin{equation}
\label{equ:se_system_model_channel_signal_channel}
h_{u}[v]  = \sum_{l = 0}^{ L - 1} p[l] g_{u}[l] e^{-j2\pi l v/N}.
\end{equation}
where $p[l]$ is the power delay profile of the channel $l-$th tap with $\sum_{l = 0}^{L -1}p[l] = 1$, $N$ is the number of channel uses, $g_{u} \sim \mathcal{C}\mathcal{N}(0,1)$, is the complex channel gain from the transmitter to the $u-$th receive antenna of the $l-$th path.

To allow for analysis, we assume  $\boldsymbol{h}[v] \approx \boldsymbol{h}[v-1]$, hence the received signal can be written as 
\begin{equation}
\begin{aligned}
\label{equ:se_II_Rx_Signal_1_16DAPSK}
y_{u}[v]     &=  \bigg(\frac{a[v]}{a[v-1]} \bigg) y_{u}[v - 1] s[v] + {z}^{'}_{u}[v^{}],\\
y_{u}[v]  &=  a^{'}[v] y_{u}[v - 1] s[v] + {z}^{'}_{u}[v^{}],
\end{aligned}
\end{equation}
where $ {z}^{'}_{u}[v^{}] =  {z}_{u}[v^{}] -  a^{'}[v]  s[v]  {z}_{u}[v^{} - 1]$.  Note that ${z}^{'}_{u}[v^{}] \sim \mathcal{C}\mathcal{N}(0,\varrho_z)$, where $\varrho_z = 2\sigma_z^2$ if $b_{1}[v] = 0$. Likewise, if $b_{1}[v] = 1$, then $\varrho_z = \sigma_z^2(1 + \frac{\psi_0^2}{\psi_1^2})$,  or $\varrho_z = \sigma_z^2(1 + \frac{\psi_1^2}{\psi_0^2})$. The signal-to-noise ratio translates to $\rho =\frac{1}{\varrho_z}$, and we define the amplitude ratio as $a^{'}[v] = \frac{a[v]}{a[v - 1]}$. The received signal across all antennas can be written in the vectorized form
\begin{equation}
\begin{aligned}
\label{equ:se_II_Rx_Signal_1_16DAPSK_vector}
\boldsymbol{y}[v]     =  a^{'}[v] \boldsymbol{y}[v - 1] s[v] + \boldsymbol{z}^{'}[v^{}].
\end{aligned}
\end{equation}
\section{Receive Processing}
At the base station, each of the receive antennas is equipped with a low-resolution ADC which is specified as a quantizing function. The quantizing function is defined as $q = \mathcal{Q}(s)$, where $\mathcal{Q}: \mathbb{C} \rightarrow \mathcal{A}_c$, and $ \mathcal{A}_c$ is the set of quantization alphabets. The quantizer independently compares the real and imaginary part of the received signal to a set of predefined thresholds. The thresholds are defined as:
$$
-\infty = \zeta_0 < \zeta_1 < \cdots < \zeta_{Q} < \zeta_{Q+
1} = + \infty.
$$
The label $q$ is unique across different quantization bins. To concisely represent the information from the quantization function we need $2\lceil\log_2Q \rceil$ number of bits.
\subsection{One-Bit Receive Processing}
In this section, we consider the one-bit quantizer defined with the $\sign $ function. The quantized version of the received signal at the $u$th antenna can be defined as
\begin{equation}
\label{equ:section_II_quantized_received_signal_1}
q_{u}[v] = \sign(\Re({y}_{u}[v])) + j\sign(\Im({y}_{u}[v]))),
\end{equation}
The quantized symbols across all antennas at the $v$th channel use can be compactly written as
\begin{equation}
\label{equ:section_II_quantized_received_signal_2}
{\boldsymbol{q}}[v] =  [q_{1}[v], q_{2}[v], \cdots, q_{U}[v] ].
\end{equation} 

To find an approximation, we use the Bussgang Theorem \cite{9307295} to decompose the received quantized signal to the unquantized signal and an uncorrelated noise
\begin{equation}
\label{equ:section_II_real_likelihood_2}
\begin{aligned}
{\boldsymbol{q}}[v] &= \eta_v {\boldsymbol{y}}[v] + \epsilon_v, \\
\end{aligned}
\end{equation}
where the $\eta$ is the quantization  scaling factor  and $\epsilon \sim \mathcal{C}\mathcal{N}(0, \sigma^2_{\epsilon})$ is the quantization noise component which is uncorrelated to $\eta$.  With some algebraic manipulation, we write (\ref{equ:se_II_Rx_Signal_1_16DAPSK_vector}) as
\begin{equation}
\label{equ:se_II_Rx_Signal_1_16DAPSK_1}
\frac{1}{\eta_{v^{} }} {q}_u[v^{}] - \frac{\epsilon_{{v^{}} }}{\eta_{v^{} }}= a^{'}[v] s[v] \bigg[ \frac{1}{\eta_{v^{} - 1}} {q}_u[v^{} - 1] - \frac{\epsilon_{v^{} - 1}}{\eta_{v^{} -1}} \bigg]  + {z}^{'}_{u}[v^{}]^{}.
\end{equation}
Assuming that the quantization effects remain constant during the entire transmission such that $\eta_v = \eta_{v - 1} = \eta$ and $\epsilon_v = \epsilon_{v - 1} = \epsilon$, we write
\begin{equation}
\label{equ:se_II_Rx_Signal_1_16DAPSK_2}
\begin{aligned}
 {q}_u[v^{}] &= a^{'}[v]  s[v] {q}_u[v^{} - 1]   - a^{'}[v]  s[v] \epsilon_{{ } }  + \epsilon_{{} } + \eta_{} {z}^{'}_{u}[v^{}] ,   \\  
 {q}_u[v^{'}] &= a^{'}[v]  s[v] {q}_u[v^{} - 1]   +  {w}^{}_{u}[v^{}],
 \end{aligned}
\end{equation}
where ${w}^{}_{u}[v^{}]  =    \eta_{} {z}^{'}_{u}[v^{}] -  a^{'}[v]  s[v] \epsilon_{{} }  + \epsilon_{{} }    $ is the combined effect of the thermal noise and the quantization noise. Note that ${z}^{'}_{u}[v^{}] \sim \mathcal{C}\mathcal{N}(0,\varrho_{z,\epsilon})$, where $\varrho_{z,\epsilon} =  \eta_{}^{2}\varrho_{z} + 2\sigma^2_{\epsilon} $ if $b_{1}[v] = 0$. Likewise, if $b_{1}[v] = 1$, then $\varrho_{z,\epsilon} = \eta_{}^{2}\varrho_{z} + \sigma_{\epsilon}^2(1 + \frac{\psi_0^2}{\psi_1^2})$,  or $\varrho_{z,\epsilon} = \eta_{}^{2}\varrho_{z} + \sigma_{\epsilon}^2(1 + \frac{\psi_1^2}{\psi_0^2})$. The signal-to-noise ratio translates to $\rho =\frac{1}{\varrho_{z,\epsilon}^2}$.

For ease of analysis, we transform the system from the complex domain to the real domain. First, the quantized and unquantized received signal can be written as

\begin{equation}
\label{equ:se_II_Rx_Signal_1_16DAPSK_3}
\begin{aligned}
 \boldsymbol{q}_{R,u}[v^{}] = \begin{bmatrix}
{q}_{R,u,1}[v^{}] \\
{q}_{R,u,2}[v^{}]
\end{bmatrix} = \begin{bmatrix}
\Re({q}_{u}[v^{}]) \\
\Im({q}_{u}[v^{}]) \end{bmatrix}, 
\end{aligned}
\end{equation}

\begin{equation}
\label{equ:se_II_Rx_Signal_1_16DAPSK_3a_1}
\begin{aligned}
 \boldsymbol{y}_{R,u}[v^{}] = \begin{bmatrix}
{y}_{R,u,1}[v^{}] \\
{y}_{R,u,2}[v^{}]
\end{bmatrix} = \begin{bmatrix}
\Re({y}_{u}[v^{}]) \\
\Im({y}_{u}[v^{}]) \end{bmatrix}, 
\end{aligned}
\end{equation}
the channel between the transmitter and the $u-$th receiver, during the $v-$th channel use can be written as
\begin{equation}
\label{equ:real_channel}
\begin{aligned}
\boldsymbol{H}_{R,u}[v] = \begin{bmatrix}
\Re({{h}}_{u} [v]) & \Im({{h}}_{u} [v])\\
 -\Im({{h}}_{u} [v]
) & \Re({{h}}_{u} [v]) 
\end{bmatrix}^{T} = \begin{bmatrix}
\boldsymbol{h}_{R,u,1}^{T}[v] \\
\boldsymbol{h}_{R,u,2}^{T}[v]
\end{bmatrix} \in \mathbb{R}^{2 \times 2}.
\end{aligned}
\end{equation}
Next, the quantized signal at the previous symbol interval received at the $u-$th base station antenna is converted from the complex to the real domain

\begin{equation}
\label{equ:se_II_Rx_Signal_1_16DAPSK_4}
\begin{aligned}
\boldsymbol{F}_{R,u}[v] &= \begin{bmatrix}
\Re({q}_u[v^{} - 1]) & \Im({q}_u[v^{} - 1])\\
 -\Im({q}_u[v^{} - 1]) & \Re({q}_u[v^{} - 1]) 
\end{bmatrix}^{T} \\
&= \begin{bmatrix}
\boldsymbol{f}_{R,u,1}^{T}[v] \\
\boldsymbol{f}_{R,u,2}^{T}[v]
\end{bmatrix} \in \mathbb{R}^{2 \times 2}.
\end{aligned}
\end{equation}
To ensure a compact derivation of the likelihood of the received signal, we enforce a sign change on both the quantized receivd signal
 $\boldsymbol{f}_{u}[v]$ as
\begin{equation}
\label{equ:se_II_sbse_unquantized_refinement_1}
\widetilde{\boldsymbol{F}}_{R,u}[v] = \begin{bmatrix}
\widetilde{\boldsymbol{f}}_{R,u,1}^{T}[v] \\
\widetilde{\boldsymbol{f}}_{R,u,2}^{T}[v]
\end{bmatrix},
\end{equation}
where $\widetilde{\boldsymbol{f}}_{R,u,i}^{T}[v]  $ is defined as
\begin{equation}
\label{equ:se_II_sbse_unquantized_refinement_2}
\widetilde{\boldsymbol{f}}_{R,u,i}^{T}[v]  =  {q}_u[v^{}]  {\boldsymbol{f}}_{R,u,i}^{T}[v].
\end{equation}
The noise and the transmit signal can be written as
\begin{equation*}
\begin{aligned}
\boldsymbol{w}_{R,u}[v] &= \begin{bmatrix}
\Re({w}_{u}[v]) \\
 \Im({w}_{u}[v])
\end{bmatrix} = \begin{bmatrix}
{w}_{R,u,1}[v] \\
{w}_{R,u,2}[v] 
\end{bmatrix} \in  \mathbb{R}^{2 \times 1} \\
  {\boldsymbol{s}}_{R}[v^{}] &= \begin{bmatrix}
\Re({{s}}[v^{}]) \\
 \Im({{s}}[v^{}])
\end{bmatrix} \in  \mathbb{R}^{2 \times 1}.
\end{aligned}
\end{equation*}
Finally, the quantized received signal can be written as
\begin{equation}
\label{equ:se_II_sbse_real_signal_1_16DAPSK}
\begin{aligned}
 \boldsymbol{q}_{R,u}[v^{}] =  a^{'}[v] \widetilde{\boldsymbol{F}}_{R,u}[v] {\boldsymbol{s}}_{R}[v^{}] + \boldsymbol{w}_{R,u}[v]  ,
\end{aligned}
\end{equation}
and based on the value of ${q}_{R,u,i}[v^{}]$, we define two sets of indices, $\mathcal{P}$ and $\mathcal{N}$ 
$$
\mathcal{P} = \{(i,u) : {q}_{R,u,i}[v^{}]  \geq 0\}, \; \; \mathcal{N} = \{(i,u) : {q}_{R,u,i}[v^{}]  < 0\}. 
$$
Finally, with these definitions, we can write the likelihood function as 
\begin{equation}
\label{equ:section_II_real_likelihood_1}
\begin{aligned}
&L(a^{'}[v] {\boldsymbol{s}}_{R}[v^{}]|\rho)  = \\
&Pr \Bigg(a_{}^{'}[v]\sqrt{\rho}{\Tilde{{\boldsymbol{f}}}_{R,u,i}^{T}[v] {\boldsymbol{s}}_{R}[v^{}]} \geq -{w}_{R,u,i}[v]  \Bigg| \forall (i,u) \in \mathcal{P} \Bigg) \\
&Pr \Bigg(a_{}^{'} [v]\sqrt{\rho}{\Tilde{{\boldsymbol{f}}}_{R,u,i}^{T}[v] {\boldsymbol{s}}_{R}[v^{}]}  \geq {w}_{R,u,i}[v] \Bigg| \forall (i,u) \in \mathcal{N} \Bigg)\\
 \equA & \prod_{i=1}^{2}  \prod_{u=1}^{U} \Phi \Bigg( a_{}^{'}[v]\sqrt{\rho} \Tilde{{\boldsymbol{f}}}_{R,u,i}^{T}[v] {\boldsymbol{s}}_{R}[v^{}] \Bigg) 
\end{aligned}
\end{equation}
Using (\ref{equ:section_II_real_likelihood_1}), the maximum likelihood detector can be written as
\begin{equation}
\label{equ:section_II_real_maximum_likelihood_1} 
\begin{aligned}
\hat{a}^{'}[v] \hat{\boldsymbol{s}}_{R}[v] = \argmax_{(a^{'}_{t})^{'} \boldsymbol{s}_{R}^{'} [v]\in \{ \mathcal{A} \times S_R \}}  
L((a^{'})^{'}[v] {\boldsymbol{s}}^{'}_{R}[v^{}]|\rho).
\end{aligned}
\end{equation}

The bit $b_{1}[v]$ can be recovered using 
\begin{equation}
\label{equ:se_II_ss_B_bit_ML_ZF}
 \hat{b}_{1}[v] = \begin{cases}
      0 , & \text{if $ \norm{\hat{a}^{'}[v] {\boldsymbol{\hat{s}}}_{R}^{}[v^{}] } = 1$},\\
      \\
     1 ,  & \text{if $\norm{\hat{a}^{'}[v] {\boldsymbol{\hat{s}}}_{R}^{}[v^{}] } \neq 1$},\\
    \end{cases} \\
\end{equation}
and with an abuse of notation the remaining bits can be recovered from the phase information by $ \{\hat{b}_{2}[v], \hat{b}_{3}[v],\cdots, \hat{b}_{N_b}[v] \} = \Upsilon^{-1}\bigg(\frac{\hat{a}^{'}[v] {\boldsymbol{\hat{s}}}_{R}^{}[v^{}] }{\norm{\hat{a}^{'}[v] {\boldsymbol{\hat{s}}}_{R}^{}[v^{}] }} \bigg)$.   Empirical results indicate that the detector suffers from a substantial error floor, which can be attributed to the performance of the amplitude recovery part of the detector. This is intuitive because the one-bit quantizer only represents one level of amplitude (i.e $1$ or $-1$). 
\subsection{Deep Differential Detection With One-Bit And Variable Quantization Levels}
To enable detection with one-bit, we propose to group the base station antennas such that receive antennas in the same group have the same quantization levels and receive antennas in different groups employ different quantization levels. In this article, the following quantization groups are used $\mathcal{U}_1,\mathcal{U}_2$, $\mathcal{U}_3$, such that $| \mathcal{U}_1| + | \mathcal{U}_2| + | \mathcal{U}_3| = U$. The quantizer operation in the $j$th group can be defined as
\begin{equation}
\label{equ:antenna_grouping_1}
\begin{aligned}
q_{R,u_j,i}[v]  =  \mathcal{Q}({{y}}_{R,u_j,i}[v]  ),
\end{aligned}
\end{equation}
where $u_j \in \mathcal{U}_j $, and ${{y}}_{R,u_j,i}[v]  \in \{\zeta_{1,j}, \zeta_{3,j}\}$. More specifically,
 $$ q_{R,u_j,i}[v]   = \begin{cases}
      \zeta_{3,j} , & \text{if $ {{y}}_{R,u_j,i}[v]  >  \zeta_{2,j}$},\\
      \\
    \zeta_{1,j} ,  & \text{if $ {{y}}_{R,u_j,i}[v]  <  \zeta_{2,j}$}.\\
    \end{cases} \\$$
Again we can decompose the received signal through the Bussgang theorem
\begin{equation}
\label{equ:reduced_precision_2bit_bussang_decomp_1b}
\begin{aligned}
\boldsymbol{{q}}_{}[v]  &=   \eta \boldsymbol{h}[v] x[v] + \eta\boldsymbol{z}[v] + {\epsilon}_{}[v], \\
\boldsymbol{{q}}_{}[v]  &=   \eta \boldsymbol{h}[v] x[v] +  \boldsymbol{\epsilon}^{'}[v],
\end{aligned}
\end{equation}
where $\boldsymbol{\epsilon}^{'}[v] = \eta\boldsymbol{z}[v] +{\epsilon}_{}[v] $ is Gaussian, i.e $\boldsymbol{\epsilon}^{'}[v] \sim \mathcal{C}\mathcal{N}(0, \Tilde{\sigma}^2_{\epsilon^{}} =  \eta^2\sigma^2_{z} + \sigma^2_{\epsilon}   )$ and considering a single antenna, $u$ in the real domain, we have
\begin{equation}
\label{equ:reduced_precision_2bit_bussang_decomp_3}
\begin{aligned}
{{q}}_{R,u,i}[v]  =  \eta  \boldsymbol{h}_{R,u,i}^{T}[v] \boldsymbol{x}_{R}[v]
  + \boldsymbol{\epsilon}_{R,u}^{'}[v], 
\end{aligned}
\end{equation}
where 
\begin{equation}
\label{equ:quant_noise_real}
\begin{aligned}
\boldsymbol{\epsilon}_{R,u}^{'}[v] =
\begin{bmatrix}
{(\epsilon_{R,u,1}^{'}[v])}\\
{(\epsilon_{R,u,2}^{'}[v])} \end{bmatrix} =
\begin{bmatrix}
\Re{(\epsilon_{R,u}^{'}[v])}\\
\Im{(\epsilon_{R,u}^{'}[v])} \end{bmatrix}, 
\end{aligned}
\end{equation}
The second order statistics of the quantized received signal at an antenna in the $j-$th group can be approximated as
\begin{equation}
\label{equ:antenna_grouping_bussang_decomp_3}
\begin{aligned}
\Lambda_i[v] &= \frac{1}{U}\sum_{u = 1}^{U} | {{q}}_{R,u_j,i}[v] |^2  \\
&=  \Tilde{x}^2[v]  \frac{\sum_{j = 1}^{3} \sum_{u \in \mathcal{U}_j}\eta^2_j\boldsymbol{h}_{R,u,i}^{H}[v]
\boldsymbol{h}_{R,u,i}[v]}{U} \\
&+ \frac{\sum_{j = 1}^{3}\sum_{u \in \mathcal{U}_j} {\epsilon_{R,u,i}^{'}[v]}^{H}
{\epsilon_{R,u,i}^{'}[v]}^{}}{U} .
\end{aligned}
\end{equation}

Assuming that quantization gains and quantization noise variance are equal across all antenna groups, the quantized received signal can be written
\begin{equation}
\label{equ:antenna_grouping_bussang_decomp_4}
\begin{aligned}
\Lambda_i[v] = \frac{1}{U}\sum_{u = 1}^{U} | {{q}}_{R,u,i}[v] |^2 =  \Tilde{x}^2[v] \eta^2 \alpha^2_i + \Tilde{\sigma}_{\epsilon}^2 .
\end{aligned}
\end{equation}

Note  due to channel hardening \cite{7880691}, as $U \rightarrow \infty$,  $ \frac{\sum_{u=1}^{U}\boldsymbol{h}_{R,u,i}^{H}[v]\boldsymbol{h}_{R,u,i}[v]}{U} $ converges to a constant, $\alpha_i$. A maximum likelihood detection approach  based on the observation $\Lambda$  is used to test the hypothesis that the symbol amplitude remains constant across adjacent symbols i.e, $\Tilde{x}[v] = \Tilde{x}[v - 1]$. More specifically,  a hypothesis testing rule can be used to determine if $b_1[v] = 1$ or if $b_1[v] = 0$. This hypothesis is defined as  $\mathcal{H}_1$ and is confirmed if 
\begin{equation}
\label{equ:conditiona_lambda_1}
\begin{aligned}
\Omega(\Lambda|\mathcal{H}_1) > \Omega(\Lambda|\mathcal{H}_0) ,
\end{aligned}
\end{equation}
where $\Omega(\Lambda|\mathcal{H}_1)$ is the conditional pdf of $\Lambda[v]$. Hence, the hypothesis test in (\ref{equ:conditiona_lambda_1}) is used to develop an energy detection threshold between two neighbouring DAPSK concentric circles.

Assuming that the channel amplitude, the quantization effect, and composite noise variance are known, the conditional pdf of $\Lambda$ follows a non-central chi-square distribution and can be written as
\begin{equation}
\label{equ:pdf_lambda}
\begin{aligned}
\Omega(\Lambda| \alpha, \Tilde{x}[v],\eta, \Tilde{\sigma}_{\epsilon}^2) &= \frac{U}{\Tilde{\sigma}_{\epsilon}^2} \bigg(\frac{\Lambda}{\alpha^2 \Tilde{x}^2[v] \eta^2} \bigg)^{U - 1} e^{-\frac{U}{\Tilde{\sigma}_{\epsilon}^2} (\Lambda +  \alpha^2\Tilde{x}^2[v] \eta^2)} \\
&\boldsymbol{I}_{U - 1} \bigg( \frac{2 U}{\Tilde{\sigma}_{\epsilon}^2} \sqrt{\Lambda \alpha^2\Tilde{x}^2[v] \eta^2}  \bigg),
\end{aligned}
\end{equation}
where $U > 0$ and $\boldsymbol{I}_{U - 1}$ is the modified Bessel function of the first kind. While the distribution is dependent on the amplitude of the transmitted symbol, $\Tilde{x}[v] \in \{\psi_0, \psi_1\}$, the respective distributions conditioned on either $\psi_0$ or $\psi_1$ are not symmetric, therefore, no closed form solution for hypothesis testing can be developed. Hence, we employ a neural network to select the correct hypothesis.

Because  $\boldsymbol{h}[v] \approx \mathbf{h}[v-1]$, the second order statistics at the current and previous channel uses  denoted by $\boldsymbol{\lambda}$ are correlated. More specifically, the elements of the following vector are correlated $ \boldsymbol{\lambda}  = \{\Lambda_i[v],\Lambda_i[v-1]\}, i \in \{1,2\}$. This vector will serve as one of the inputs to the neural network based amplitude detector. 
To detect the phase information with the 1-bit detector presented in (\ref{equ:section_II_real_maximum_likelihood_1}), we allocate a particular group to use the signum quantization function. For instance, if the $j$th group is reserved for phase detection, then ${{q}}_{R,u_j,i}[v] \in \{-1, 1\}$, and $\zeta_{2,j} =0$. Hence, with an amplitude ratio $a_{}^{'}[v] = 1$, the maximum likelihood phase detector can be written as 
    \begin{equation}
\label{equ:section_II_variable_maximum_likelihood_detector} \hat{\boldsymbol{s}}_{R,VQL}[v]
=  \argmax_{{\boldsymbol{s}}_{R}^{'}[v^{}] \in \mathcal{S}_R} \prod_{i=1}^{2}  \prod_{u \in \mathcal{U}_j} \Phi\Bigg( \sqrt{\rho} \Tilde{{\boldsymbol{f}}}_{R,u,i}^{T}[v] {\boldsymbol{s}}_{R}^{'}[v^{}] \Bigg).
\end{equation}

\subsection{Structure of Neural Network Based Amplitude Detector}
The NN-based amplitude detector is represented by a DNN with $L$ fully-connected (FC) layers. The operation of the $l$th layer of the neural network can be described as 
\begin{equation}
\label{equ:ch2_f^l}
\mathbf{w}^{l} = f_{\theta^{l}}(\mathbf{w}^{l-1}) = \varphi^{l}(\mathbf{W}^{l} \mathbf{w}^{l-1} + \mathbf{\beta}^{l} ),
\end{equation}
where $\mathbf{W}^{l}$ and $\mathbf{\beta}^{l}$ describes the weights and biases terms of the $l$th layer, $\mathbf{w}^{l-1}$ describes the output of the previous layer,  $\varphi^{l}$ denotes the activation function of the $l$th layer, and $\mathbf{\theta}^{l} = \{ \mathbf{W}^{l} , \mathbf{\beta}^{l} \}$ denotes the parameters of the $l$th layer. The NN amplitude detector can be described as $\theta =\{\mathbf{\theta}^{1}, \mathbf{\theta}^{2},\cdots, \mathbf{\theta}^{L}\}$. Note that at the 1st layer, $\mathbf{w}^{0}$ denotes the input to the NN-based decoder. This input consists of the second order statistics represented by $\boldsymbol{\lambda} $ concatenated with a one-hot representation of the measured SNR. The one-hot represenation of the SNR is denoted as $\boldsymbol{\varrho}$. Hence, $\mathbf{w}^{0} = \{\boldsymbol{\lambda},  \boldsymbol{\varrho}\}$. The first $L-1$ layers are equipped with a Relu activation function \cite{agarap2018deep}, while the $L$th layer in the neural based decoder employs the Softmax function \cite{bridle1990training}. The output of the Softmax layer denoted by $\mathbf{w}^{L}$ provides two pseudo-probabilities, each denoting the likelihood that the decoded label is either zero or one. Hence, the operation of bit prediction can be described as
\begin{equation}
    \label{equ:bits_recovery}
 \hat{b}_{1}[v] = \begin{cases}
        0 , & \text{if $ \arg\max_{i} {\mathbf{w}^{L}[i]} = 1$ },\\

     1 ,  & \text{else if $ \arg\max_{i} {\mathbf{w}^{L}[i]} = 2$}.\\
    \end{cases} \
\end{equation}

In summary, the NN-based amplitude detector operates by comparing two signals received during adjacent channel uses and determines whether the transmitted symbols are both from the inner constellation or from the outer constellation. If the neural network determines that the transmitted symbols are from the same circle, the prediction is $\hat{b}_{1}[v] = 0$ , otherwise the prediction is $\hat{b}_{1}[v] = 1$.
\subsection{Training Procedure of Neural Network Based Amplitude Detector}
The NN-based amplitude detector is trained offline using a randomly generated dataset.  The dataset consists of a collection of blocks of bits, $\{\boldsymbol{b}[v]\}_{v=1}^{V}$, and a corresponding collection of differential modulated symbols, $\{{x}[v]\}_{v=1}^{V}$. The left-most bit in each block of bits is converted to one-hot encoded vectors with two elements denoted as $\{\boldsymbol{\omega}[v]\}_{v =1}^{V}$. At each channel use, fading channel vectors and random Gaussian noise vectors are generated. The measured SNR is converted to one-hot encoded vectors. We adopt the binary cross entropy loss function for training
\begin{equation}
\label{equ:trianing_loss_fn}
\mathcal{L}(\mathbf{\theta}) =  - \frac{1}{V}\sum_{v = 1}^{V} \boldsymbol{\omega}^{T}[v] \log({\mathbf{w}^{L}} ),
\end{equation}
equipped with this loss function and a learning rate, $\alpha$, the stochastic gradient descent algorithm is used to update the neural network parameters
\begin{equation}
\label{equ:trianing_loss_sgd}
\mathbf{\theta} := \mathbf{\theta} - \alpha\nabla\mathcal{L}(\mathbf{\theta}).
\end{equation}
In this work, an advanced version of the stochastic gradient descent algorithm - the Adam optimizer \cite{kingma2014adam} is used to update the neural network parameters. The training parameters are presented in Table \ref{table:table_training_params}.  The test dataset is similarly generated. The instantaneous values of test dataset is different from the training dataset, but have the same statistics.
\begin{table}[!t]
\renewcommand{\arraystretch}{1.3}
\caption{Training Parameters}
\label{table:table_training_params}
\centering
\begin{tabular}{|c||c|}
\hline
Learning Rate, $\alpha$& $0.001$    \\
\hline
Iterations (epochs)& $250$  \\
\hline
Batch size, $V$& $1000$   \\
\hline
\end{tabular}
\end{table}
\section{NUMERICAL RESULTS}
We perform Monte-Carlo simulations to evaluate the proposed differential amplitude and differential phase detection schemes. The proposed differential schemes is compared with corresponding coherent schemes presented in \cite{7088639,7094619}. We consider $N = 256$ channel uses. In the coherent scheme, a  fraction of channel uses, $\xi$, is used for the transmission of pilot symbols. The differential scheme use no pilot symbols. The variable quantization level setup consist of three groups with equal number of antennas $| \mathcal{U}_1| = | \mathcal{U}_2| = | \mathcal{U}_3| = \frac{U}{3}$. The second group is used for phase detection and each antenna in this group employs the signum function for quantization. The first and third groups employ one-bit quantization with the following thresholds, $\zeta_{2,1} = \frac{\psi_0(1 + a \cos{\frac{\pi}{4}})}{2} $ and $\zeta_{2,3} = -\zeta_{2,1}$. The output of the quantizers used in the first group is described as
 $$ {{q}}_{R,u_1,i}[v]  = \begin{cases}
      \psi_1 , & \text{if $ {{y}}_{R,u_1,i}[v] >  \zeta_{2,1}$},\\
      \\
    \psi_0 ,  & \text{if $ {{y}}_{R,u_1,i}[v] <  \zeta_{2,1}$}.\\
    \end{cases} \\$$ and  the output of the quantizers used in the second group is specified as
    
     $$ {{q}}_{R,u_3,i}[v]  = \begin{cases}
      -\psi_0 , & \text{if $ {{y}}_{R,u_3,i}[v] >  \zeta_{2,3}$},\\
      \\
    -\psi_1 ,  & \text{if $ {{y}}_{R,u_3,i}[v] <  \zeta_{2,3}$}.\\
    \end{cases} \\$$
    
    \begin{figure}[!htb]
\centering
\includegraphics[width=\linewidth]{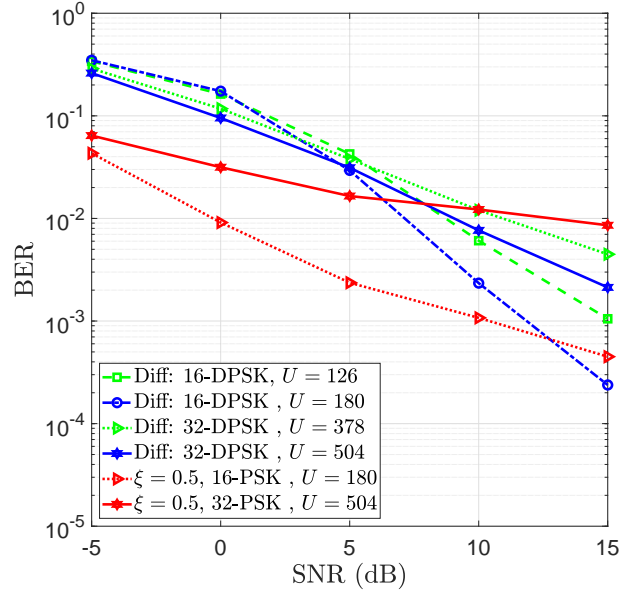}
\caption{ BER for DPSK in the differential system in comparison with the BER for PSK in the coherent scheme. The channel used in this simulation is frequency selective with $L = 31$ channel taps. \emph{$\xi$ indicates the fraction of channel uses need for pilot transmission in the coherent system. "Diff" indicates differential modulation}. } 
\label{fig:Results/ber}
\end{figure}

    \begin{figure}[!htb]
\centering
\includegraphics[width=\linewidth]{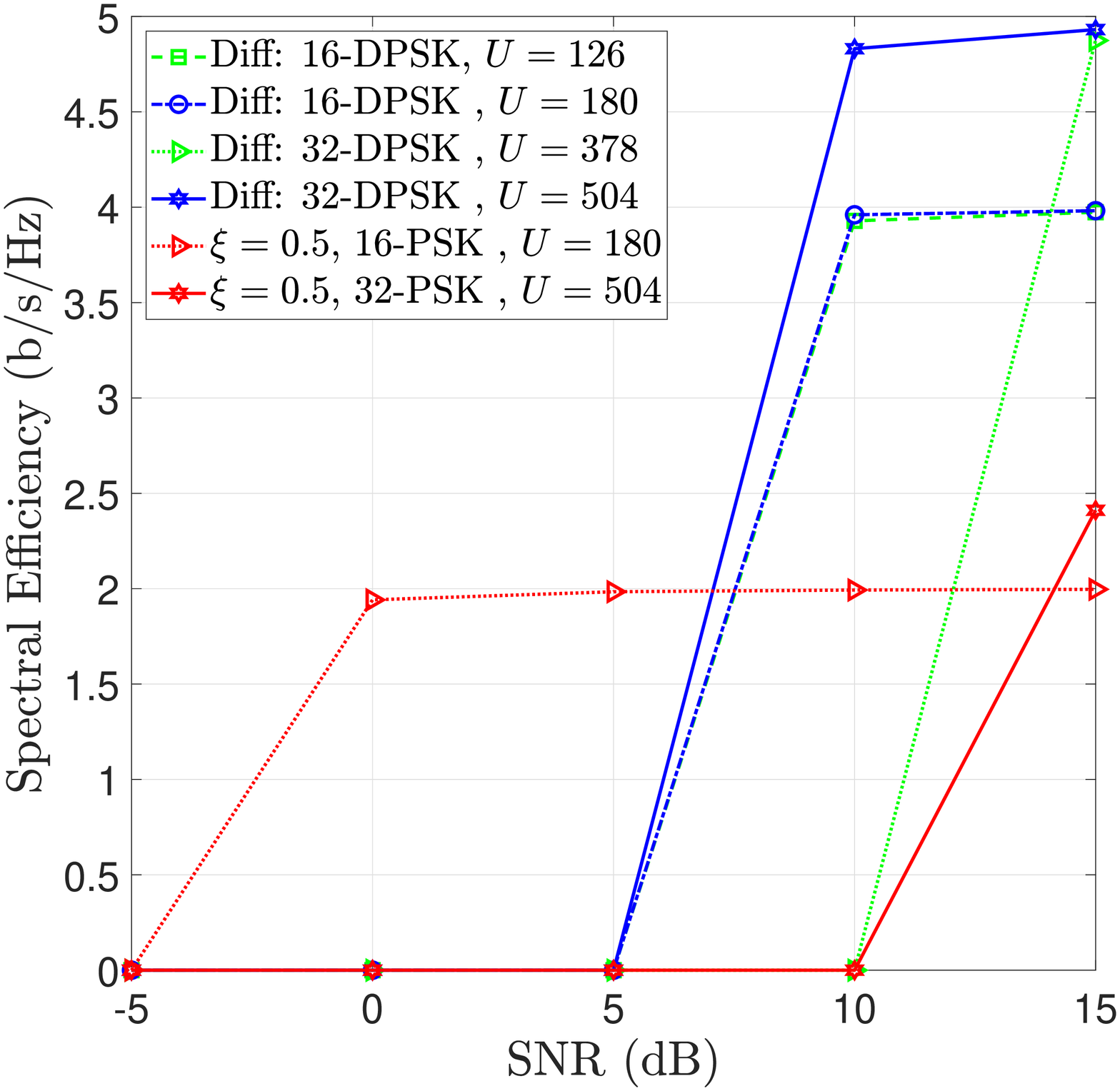}
\caption{ Spectral efficency attained in the differential system in comparison with the spectral efficency attained using PSK in the coherent scheme. The channel used in this simulation is frequency selective with $L = 31$ channel taps.  \emph{$\xi$ indicates the fraction of channel uses need for pilot transmission in the coherent system. "Diff" indicates differential modulation}. } 
\label{fig:Results/spe}
\end{figure}
Figure \ref{fig:Results/ber} presents the BER incurred in decoding the block of information bits - $\boldsymbol{b}[v] = [b_{1}[v], b_{2}[v], b_{3}[v],\cdots, b_{N_b}[v]]$. The amplitude information (first bit) is decoded with the neural network as presented in (\ref{equ:bits_recovery}). The phase information symbol (remaining $N_b - 1$ bits) is decoded with (\ref{equ:section_II_variable_maximum_likelihood_detector}) and $ \{\hat{b}_{2}[v], \hat{b}_{3}[v],\cdots, \hat{b}_{N_b}[v] \} = \Upsilon^{-1}\bigg(\frac{ {\boldsymbol{s}}_{R,VQL}^{'}[v^{}] }{\norm{ {\boldsymbol{s}}_{R,VQL}^{'}[v^{}] }} \bigg)$.

The BER attained for both the differential and coherent detectors  decreases with an increase in receive antennas and with an increase in signal-to-noise ratio. The coherent system employs $\xi = 0.5$ of the total channel uses for transmitting pilot symbols. The BER performance of coherent 16-PSK is better than the BER performance of the differential 16-DPSK. The performance gap between the differential system and the coherent system is reduced when the modulation is increased from $16$ to $32$. 

Figure \ref{fig:Results/spe} presents the spectral efficiency attained in both the differential and the coherent system. This spectral efficency is calculated with
$$
S.E.   = \begin{cases}
      \xi N N_b (1-  SER) , & \text{if $ SER \geq SER_{th}$},\\
    0,  & \text{otherwise},\\
    \end{cases} \\$$
where $SER_{th}$ is a threshold of the symbol error rate which is derived from the block error rate. In this work, this $SER_{th}$ value is set to $5\%$. In the spectral efficiency plot, the advantage of the differential system is apparent. For a modulation order of 16, the coherent system has a better BER, but the spectral efficiency of the differential system is  two-times better than the spectral efficiency of the  coherent system. This is because the one-bit coherent system uses more than half of the available channel uses for pilot symbols. Note that at high SNR, the spectral efficiency of the differential system is much higher than the spectral efficiency of the coherent system, irregardless of modulation order.
\section{Conclusion}
This article has investigated the uplink of a massive MIMO system with differential amplitude and phase modulation employed at the transmitter and one-bit ADCs employed at each receive antenna at the base station. The Bussgang theorem is used to express the quantized received signal in terms of quantized signals received during previous channel uses. With this expression, we derived the maximum likelihood expression for the differential encoded  amplitude and phase information symbols. Because the maximum likelihood detector failed to decode the amplitude information, we  developed and trained a neural network-based amplitude detector. We validated the performance of the proposed detectors through Monte-Carlo simulations and provided a comparison with coherent one-bit detectors. Our results indicate that the one-bit differential system outperforms the one-bit coherent system in terms of spectral efficiency.






%


	\bibliography{refs}
	\bibliographystyle{IEEEtran}

\end{document}